**Digital health transformation in Quebec: assessment of interoperability and governance strategies**

Alexandra Langford-Avelar[1,2], Delphine Bosson-Rieutort[1,2,3]

[1] Département de gestion, évaluation et politique de santé, École de santé publique de l'Université de Montréal (ESPUM), 7101 Park Ave, Montreal, Quebec, Canada, H3N 1X9

2 Centre de recherche en santé publique (CReSP), Université de Montréal and Centre intégré universitaire de santé et de services sociaux du Centre-Sud-de-l'Île-de-Montréal, Montréal, Québec, Canada, 7101 Park Ave, Montreal, Quebec, Canada, H3N 1X9

[3] Centre interuniversitaire de recherche en analyse des organisations (CIRANO), 1130 Rue Sherbrooke O #1400, Montréal, Quebec, Canada, H3A 2M8

**Corresponding authors**

Alexandra Langford-Avelar, 7101 Park Ave, Montreal, Quebec, Canada, H3N 1X9, +1(514-715-1494)  alexandra.langford.avelar@umontreal.ca

**CRediT authorship contribution statement**

**Alexandra Langford-Avelar:** Conceptualization, Data curation, Formal analysis, Investigation, Writing – original draft. **Delphine Bosson-Rieutort:** Supervision, Writing – translation, review & editing.

*This manuscript was initially written in December 2023 and subsequently updated in December 2025*

**Abstract**

The rapid expansion of health data has led to unprecedented information availability within healthcare systems. Health information systems (HIS) play a central role in managing this data and enabling improvements in care delivery, system performance, and population health monitoring. Maximizing the value of HIS, however, requires effective information exchange across systems, making interoperability a critical prerequisite. Despite its recognized benefits, interoperability remains a major challenge within Quebec's Health and Social Services Network, largely due to the heterogeneity and fragmentation of HIS across healthcare institutions.

This paper assessed how Quebec's *Plan sante* addressed interoperability challenges, using the 4 dimensions from the Healthcare Information and Management Systems Society (HIMSS): foundational, structural, semantic, and organizational interoperability. This study highlighted initiatives aimed at strengthening infrastructure and information system architecture to support foundational interoperability and showed persistent challenges at the structural and semantic levels, particularly those related to the adoption of standardized data formats and harmonization of clinical terminologies. Finally, significant implementation challenges that require coordinated change management were identified regarding the organizational interoperability.

Overall, while the *Plan sante* demonstrates a clear commitment to technological modernization, it does not fully address the interoperability multidimensional nature. Achieving meaningful interoperability will require sustained efforts across technical, normative, and organizational domains beyond the strategies currently outlined. Recent governance developments, including the creation of *Sante Québec*, add complexity to this evolving context and raise further questions regarding the coordination of interoperability governance.

**Background and conceptual framework**

The growing importance of health data has transformed our health systems. With a rapid technological progress and increasing digitization, the volume of available health information has expanded considerably, emerging from diverse sources such as electronic medical records (EMR), laboratory systems, imaging and other connected medical devices. When properly structured and processed, these data have the potential to improve the quality of care, promote research and optimize the overall management of the healthcare systems [1]. The health information systems (HIS) play a central role in the management and exploitation of this information richness, paving the way for significant advances in healthcare delivery. As integrated ensembles of components, they enable the collection, storage, processing and dissemination of relevant information to support the management of an organization [1]. Within our healthcare organizations, various HIS are deployed to support clinical decision-making, facilitate the coordination of care, promote medical research and allow for performance monitoring, human resources management and resources planification [1,2]. Their effectiveness, however, depends heavily on their capacity to share information between various system, such as transmitting X-ray results to the hospital information system, while preserving the accuracy and meaning of the shared information. The key purpose is to simplify access and retrieval of clinical data, thereby promoting supported, timely and effective care delivery.

In this context, interoperability, as defined by the Healthcare Information and Management Systems Society (HIMSS), a recognized global advisor in digital health transformation, refers to the ability of information systems to access, exchange, integrate and collaboratively use data, both within and across organizational, regional and national boundaries [3]. The interoperability has become essential to the performance of modern health systems to enable the smooth and rapid transmission of information from one information system to another. Without it, data remain isolated, within institutional silos, impeding the generation of timely and representative outcomes, leading to an incomplete understanding of individual and population health utilization and need, resulting in suboptimal care delivery and increased healthcare costs [4]. Interoperability, as conceptualized by HIMSS, is achieved through four levels [3]. Foundational interoperability refers to the basic capacity of systems to exchange data securely without standardization or interpretation. Structural interoperability introduces syntactic consistency, allowing data to be transmitted in predictable formats that can be processed by receiving systems. The data is structured in a standard way, like a sentence composed of a pronoun, a verb, and a complement. The most common structural standard in healthcare is HL7 [4]. Semantic interoperability ensures that exchanged information retains a consistent meaning across platforms using harmonized and recognized terminologies and classifications such as SNOMED CT, ICD, or LOINC. Finally, organizational interoperability encompasses the governance structures, regulatory frameworks, institutional arrangements, and workflows required to enable secured, coordinated and responsible data use. Hence, interoperability is not just about the ability to exchange information; it also requires the ability to understand the data being exchanged.

Despite the recognized advantages of interoperable HIS, the situation in Quebec, Canada, remains challenging. In 2019, the Quebec health and social services network (*Réseau de la santé et des services sociaux*, RSSS), contained more than 9,000 HIS and at least 400,000 users [5]. These HIS vary significantly from one healthcare facility to another, as they are mainly tailored for local access, resulting in significant heterogeneity in design, functionality and standards. As such, patient information may be scattered across various institutions, and data generated in one facility are frequently inaccessible in another. Even worse, within a same facility, some technologies may not communicate with others, forcing clinicians to rely on outdated technologies such as fax machines, undermining timeliness, completeness and accuracy of information needed for safe and individual care, at the expense of the patients' health [5]. The COVID-19 pandemic has unfortunately greatly highlighted the consequences of this fragmentation and the urgent need for a fluid and reliable exchange of information within the Quebec health services and social network. Facing these challenges, a substantial upgrade of information technology is essential. Recognizing the persistence of significant technological delays, the government launched an ambitious digital transformation agenda through the "Plan to implement necessary changes in healthcare" *(Plan pour mettre en œuvre les changements nécessaires en santé,* or *"Plan santé")* [6] which includes the 2022-2025 Technology Modernization Plan *(Plan de modernisation technologique 2022-2025)* [7]. However, despite the government's commitment to improve information exchange, questions remain regarding the ability of the *Plan santé* to address the challenges related to interoperability within the Quebec health and social services network. The outlined guidelines seem to be insufficient to adequately resolve interoperability issues.

To explore these concerns, this analysis draws on the four dimensions of interoperability defined by the HIMSS as: foundational, structural, semantic, and organizational. By examining the orientations of the *Plan santé* [6], the technological modernization plan [7], and the associated digital strategies through this framework, we assessed whether Quebec's proposed approach adequately addresses the technical, normative, and governance elements required to achieve comprehensive interoperability across its health system.

**Observation based on the HIMMS framework**

*Foundational interoperability*

To establish foundational interoperability (level 1), it is imperative to meet essential requirements. The technological infrastructure including the various components of the system, must be sufficiently adapted to enable information sharing, and optimizing the architecture of information systems is crucial to promote efficient exchange [8].

The *Plan Santé* clearly demonstrates the intention to migrate from paper-based management to a digital approach and commits to addressing the obsolescence of current technological systems by adopting modern IT solutions across various functional domains of the health network. The technology modernization plan outlined in the *Plan Santé* allocates a $292.5-million budget to upgrade technological infrastructure. The project includes objectives such as improving telecommunications network capacity, strengthening cybersecurity, consolidating data centers, and transitioning to cloud computing. This transition to a digital information system aims to overcome the inherent challenges associated with paper-based processes by introducing specialized software and optimizing the architecture of HIS to better respond to the specific requirements of each domain. This phase also aims to ensure that the necessary infrastructure is in place to support secure communication between HIS.

The *Plan Santé,* through its modernization plan, proposes clear guidelines aimed at meeting the infrastructure and architectural requirements for achieving foundational interoperability, despite the inherent challenges in their implementation. However, achieving this first level of interoperability is far from sufficient to ensure the coordinated integration and use of data from one

HIS to another. To do so, the data must be structurally and semantically standardized [9], which refers to the second and third levels of interoperability.

*Structural interoperability*

The *Plan santé* identifies the deployment of a single, province-wide Digital Health Record (*Dossier santé numérique, DSN*) as its ultimate objective for the modernization of the entire Quebec health and social services network [5]. Achieving such a unified record requires the consolidation of data generated by all HIS across the health network into a central environment, which cannot be accomplished without a shared structural foundation. In practice, this consolidation can occur at two distinct levels: *i)* within institutions, through the alignment of local systems and internal data models, and *ii)* at the regional or provincial level, through the adoption of shared standards that ensure uniform data exchange across the network. From the HIMSS perspective, both pathways rely fundamentally on achieving structural interoperability (level 2), defined as the adoption of common syntactic standards and data models to ensure consistent formatting and exchange of information across systems. For the DSN to operate as intended, the information produced by heterogeneous SIS must therefore follow province-wide structural specification [3,4]. Without structural standardization, the DSN would be forced to process inputs using incompatible data formats, undermining its reliability and scalability.

However, Quebec currently lacks formally mandated interoperability standards. Health IT vendors retain significant autonomy in selecting the technical standards on which their systems are built, and implementation are independently negotiated by each institution [10]. This decentralized approach has produced considerable heterogeneity among systems, even within identical clinical domains, resulting in a complex and various ecosystem. This structural fragmentation has been acknowledged at the federal level and the Competition Bureau of Canada's Digital Health Services Market Study from 2022 called, in its recommendation 3, for the establishment of interoperability standards for electronic health record systems and for the creation of an independent body to develop and enforce them [11]. In line with this recommendation, the Ministry of Health and Social Services (MSSS) created the *Unité de coordination des normes pour l'interopérabilité (UCNI),* mandated to develop guidelines supporting structural standardization across the health system [12]. While this initiative represents an important step toward a structural alignment, Quebec has not yet reached consensus regarding which standards should be adopted. As a result, structural interoperability remains more a technical issue delegated to vendors than an organizationally governed requirement across the health network.

*Semantic interoperability*

While structural interoperability concerns how data are formatted, semantic interoperability (level 3) ensures that data maintain consistent meaning when exchanged across systems. Within the HIMSS framework, semantic interoperability relies on harmonized terminologies, classifications, and ontologies that allow different systems to interpret clinical information in the same way. In the context of the DSN, this means ensuring that a diagnosis, laboratory test, or clinical observation conveys the same meaning regardless of the institution or system from which it originates. The *Plan santé* does not directly address this semantic layer. Although structural standardization is implicitly recognized as necessary for data consolidation, the challenge of reconciling clinical vocabularies across Quebec's health institutions remains unresolved. In principle, semantic alignment could be performed at the DSN level, allowing institutions to maintain local terminologies while mapping them to a unified semantic model at the provincial scale [9].

However, achieving this requires robust terminology-mapping mechanisms and a shared conceptual framework, none of which have yet been defined, and major challenges arise from the diversity of clinical and biomedical vocabularies used across domains. Different areas of healthcare rely on distinct terminologies, ontologies, and classification systems, often representing similar medical entities in incompatible conceptual structures. This issue is not unique to Quebec; internationally, semantic harmonization remains a recognized research challenge. For example, the U.S. National Institutes of Health (NIH) is developing a unified medical language system (UMLS) designed to link multiple biomedical semantic standards [13]. However, it is important to emphasise that this issue remains an area of ongoing research and has not yet been fully resolved [14]. The absence of consensus on how clinical concepts should be represented presents an additional barrier. Variation in ontological structures leads to inconsistent categorization, relationships, and interpretations of data. Even with consensus, technical challenges would persist, including the integration of legacy data, semi-structured information, and unstructured clinical text, which are components that constitute a significant proportion of health records and must ultimately be incorporated into the DSN [9].

Thus, whereas structural interoperability may be addressed through governance and standardization mechanisms, semantic interoperability depends on deeper conceptual alignment across biomedical domains. Neither the *Plan santé* nor the technological modernization plan currently engages with these foundational semantic challenges. As a result, even if structural interoperability were achieved, the absence of a provincial semantic framework would limit the DSN's ability to yield meaningful, interpretable, and clinically actionable information.

*Organizational interoperability*

Finally, one of the key highlights of the *Plan Santé* pertains to the restructuring of the legislative framework governing access to health data [6]. This legislative initiative aims to standardize and revise the rules surrounding the consultation and acquisition of health information, thereby ensuring an integrated management of data across the health and social services network. A preliminary step was taken in March 2021, with the adoption of the Health and Social Services Information Act [15]

Historically, in Quebec, the provisions governing access to health and social services data were fragmented across multiple laws and regulations. This fragmentation created significant complexity, hindering the work of healthcare professionals and researchers, and limiting access to comprehensive information [16]. This lack of coherence often led to delays in clinical decision-making and impeded more efficient healthcare management. In response, reforms have been introduced to streamline and consolidate the provisions related to data access, establishing a more coherent and unified framework [16].

The new law seems to specifically align with the fourth level of interoperability, namely organizational interoperability. This legislation introduces several structural changes aimed at harmonizing existing laws pertaining to data access, ensuring robust protection while simplifying the procedures for healthcare professionals. As a result, these measures support the development of integrated information management, thereby advancing the governance and data confidentiality objectives outlined in this dimension of interoperability [6,7]. However, despite the positive legislative developments, the effective implementation of these provisions within the health and social services network will require substantial effort. The legal changes will necessitate numerous adjustments to the standard operating procedures of healthcare institutions, demanding a coordinated approach to change management in order to ensure compliance and consistency across the network. The MSSS has launched an initiative in this regard, providing a new digital transformation support service in 2021. Nonetheless, the organizational challenge of digital transformation remains significant, as it often involves a cultural shift that is difficult to initiate [17].

In addition, a further challenge arises with the potential decision by the Quebec government to delegate the management of its data to external entities as part of its transition to cloud computing. This raises critical concerns regarding digital sovereignty [18]. The location of data storage infrastructure, whether within Quebec or abroad, and whether public or private, has a direct impact on an organization's ability to maintain its digital sovereignty. Some experts argue that maintaining physical control over data within the region strengthens sovereignty, whereas outsourcing data management to external providers may raise concerns about foreign regulatory frameworks [18].

**Recent governance developments in Quebec's health system**

In this evolving context, the recent creation of *Santé Québec*, a centralized agency mandated to oversee the operational performance of the health and social services network, introduces a new governance layer that may influence the implementation of digital transformation initiatives [19]. While *Santé Québec* is not explicitly framed as a digital health authority, its role in coordinating service delivery and operational priorities could indirectly affect decisions related to information systems deployment, standardization efforts, and data governance. The establishment of *Santé Québec* may be interpreted as an attempt to reinforce organizational coherence across the health system; a dimension closely aligned with the organizational level of interoperability as defined by HIMSS. By centralizing certain operational responsibilities previously distributed across institutions, *Santé Québec* could, in principle, facilitate more coordinated decision-making regarding information system alignment, change management, and the implementation of interoperability-related policies. While this structural reorganization aims to improve system-wide performance, it does not, in itself, address the foundational technical challenges associated with interoperability. At this stage, however, the articulation between *Santé Québec*'s operational mandate and existing digital health governance structures remains insufficiently documented.

**Conclusion**

To sum up, the 2022-2025 Technological modernization plan, which is part of the *Plan Santé*, aims to strengthen the technological infrastructure of the Quebec's health and social services network, thereby promoting foundational interoperability between HIS. However, challenges remain, particularly at the structural and semantic levels, due to the lack of interoperability standards, the diversity of HIS, and the complexities related to health vocabularies. The lack of consensus on structural and semantic standards makes the implementation of the DSN even more complex.

Although legislative reforms addressed organizational interoperability, questions remain regarding integrated management, privacy protection, and coordinated change management. Thus, the full achievement of interoperability requires ongoing efforts at both the technical and regulatory levels, which, do not appear to be fully addressed by the guidelines outlined in the *Plan Santé*.